\documentstyle[multicol,aps,prl]{revtex}
\begin{document}
\widetext

\draft

\title{ Onset of Wave Drag due to Generation of Capillary-Gravity Waves by a Moving Object
                   as a Critical Phenomenon}

\author {Teodor Burghelea and Victor Steinberg}

\address{Department of Physics of Complex Systems,\\
The Weizmann Institute of Science, Rehovot 76100, Israel}

\date{\today}
\maketitle
\begin{abstract}
 The onset of the {\em wave resistance}, via generation of capillary
gravity waves, of a small object moving with velocity $V$, is investigated experimentally.
Due to the existence of a minimum phase velocity $V_c$ for surface waves, the
problem is similar to the generation of rotons in superfluid helium near their
minimum. In both
 cases waves or rotons are produced at $V>V_c$ due to {\em Cherenkov
 radiation}. We find that the transition to the wave drag state is continuous: in
the vicinity of the bifurcation the wave resistance force is proportional to
$\sqrt{V-V_c}$ for various fluids.
\end{abstract}
\begin{multicols}{2}
\narrowtext

 An object, moving uniformly on a free fluid surface, is subjected to a drag force, that  can be of 
 different physical nature. The most common one is a viscous drag 
 which at low Reynolds numbers, $Re\leq 1$, is the Stokes drag, $F_v=3\pi\eta d V$, that is proportional 
 to an object velocity, $V$, and at higher $Re$ is the drag which originates from either
 laminar or turbulent wakes\cite{lan}. (Here $d$ is the sphere diameter, and $\eta$ is the viscosity.)
 However, there exists another drag force, which originates 
 from generation of capillary-gravity waves by an
 uniformly moving object. These waves remove momentum from the object into infinity, that
 produces the wave resistance force, acting on the object\cite{light}.

The dispersion relation for the surface waves is \cite{lan}
                 $$\omega^2=gk + \sigma k^3/\rho, \eqno(1) $$
where $\omega$ is the circular frequency, $k$ is the wave number,
$\rho$ is the
 fluid density, $g$ is the gravity acceleration, and
$\sigma$ is the surface
 tension. According to Eq.~(1), a phase velocity of the waves
$c=\omega/k$ has a mimimum $ V_c=({4g\sigma/\rho})^{1/4}$ at the
capillary wave
 number $k_c=\sqrt {\rho g/\sigma}$. Below $V_c$ the surface 
 waves cannot be emitted, so no 
 wave resistance force acts on the object. The stationarity 
 condition of the wave pattern in the frame moving with the object, leads to \cite{light}
 $$\omega= kV\cos\theta, \eqno (2)$$
where
$\theta$ is the angle between ${\bf V}$ and ${\bf k}$. 

Eliminating $\omega$ from equations (1) and (2), one finds
                   $$\cos{\theta (k)}=c(k)/V. \eqno(3)$$
This equation has evidently no solutions for $V<V_c$ and describes the
opening of the Cherenkov radiation cone at
$V>V_c$\cite{cher}. Generation of surface waves is analogous to the 
Cherenkov emission of electromagnetic waves by a charge uniformly 
moving with a superluminal velocity in a medium. Due to the linear dispersion
relation the  Cherenkov radiation of the electromagnetic waves occurs at 
every wave number into a cone with an opening $\theta$
 defined by Eq.~(3).

 For a given $V\geq V_c$ the system of equations (1) - (2) has solutions in
  the range  $0\geq \theta<\chi$, where the angle $\chi$ is defined by
  $\cos\chi=V_c/V$. This range of $\theta$ corresponds to the interval
  $k_1<k<k_2$ of wave numbers, resulting from the intersection of Eqs.~(1) with a straight line of
 slope $V$. 

 Therefore, unlike the electromagnetic waves, the capillary-gravity waves are
 radiated within a finite range $\Delta k=k_2-k_1$. It tends
to zero
as
 $\Delta k=4k_c\sqrt\epsilon$, when
$V$ approaches $V_c$ from above, where $\epsilon=(V-V_c)/V_c$.
 The presence of the minimum on the dispersion curve of the capillary-gravity waves (eq.(1)) results in
 a gap in the spectrum $\omega (k)$, similar to the gap in the energy spectrum of a superfluid\cite{lan1}.  
 The existence of the minimum phase velocity for the dispersion curve (1) is analogous
to the existence of the Landau critical velocity for the phonon-roton energy spectrum in a superfluid
 helium \cite{lan1}. 
 The problem of the drag onset at the Landau critical velocity $V_0$  due to the radiation
of rotons was recently considered theoretically in
 Ref.~\cite{pom}. Pomeau and Rica observed the onset of the
 roton generation at a certain critical velocity, in
numerical simulations of the generalized nonlinear Shr$\ddot{o}$dinger equation, and pointed out but without
 proof, that the roton drag force close to the onset depends on  the control parameter
 $\epsilon=(V-V_0)/V_0$ as $\sqrt \epsilon$ in 3D and as
$\log\epsilon$ in 2D.

The onset of the wave resistance and its behavior in the
 supercritical region for the capillary-gravity surface waves was
 theoretically considered in  recent papers
  \cite{raph}. The main result of the calculations is the prediction of a
  discontinuous transition to a nonzero wave resistance state at
 $V=V_c$, if the object size is much smaller than the capillary
 wavelength $\lambda_c=2\pi/k_c$\cite{raph}. \\
In this Letter we investigate the dependence of the wave drag force, on the object velocity, $V$, in 
the vicinity of the transition for various fluids and object sizes. The results of the 
experiments clearly demonstrate that the transion to the wave drag state is continuous one in 
contradiction with the theoretical prediction\cite{raph}.
  
The experiments were conducted in a circular rotating channel of 12.4 cm outer radius and 3.6 cm width, made 
of plexiglass, with a stainless steel
ball, as an object, suspended on a wire, which is glued to an elastic fiber. The channel was driven by a
stepping motor via a belt transmission with ratio 0.138 and with a 
velocity controlled better than  0.02\%. Balls of various diameters, $d$, (1.57, 2.35, and 3.14 mm) and 
wires (0.3 and 0.7 mm dia) half-immersed into a fluid, were used. Each of them is much smaller than 
$\lambda_c$ for any fluid used 
in the experiments. A drag force as a function of the velocity was measured in two ways. 
The first method was to measure the drag force, applied to an object, when the object deviates from its 
equilibrium vertical position with increasing driving velocity in accordance with the drag force. The 
second method was to measure the same force 
but with an object, hold in an initial vertical position by using an appropriate feedback loop. The latter
was 
actuated via eddy current commercial displacement gauge (EMD 1050 Electro Corporation), that measures a 
deviation of 
brass cylinder(2 mm dia), mounted on the same wire as the object, from its initial position. A signal from 
the displacement detector is used via an instrumentational amplifier and a power amplifier for the feedback
loop to drive electromagnets which hold the object in its 
initial position (see Fig.1). A current supplied to the electromagnets is calibrated as a force. In another 
setup a ball deviation from its initial position was measured 
optically by CCD camera with spatial resolution of about 10 $\mu$m. Light refraction visualization was 
also used to define the onset of the surface waves generation, the angle of the Cherenkov cone opening,
and the wavenumbers as a function of velocity. As working fluids water, silicone oils DC200 of the kinematic 
viscosities $\nu=\eta/\rho=$10 and 50
cS and of the same surface tension $\sigma=$21.2 mN/m, water-glycerol mixtures of $\nu=$18, 30, and 46 cS 
 and almost the same $\sigma=$66 mN/m, were used. The depth of each fluid was 
sufficiently large to ensure the validity of an infinite depth approximation for the surface waves.

The raw data on the full drag force, $F$, as a function of the channel velocity for silicone oil(10 cS)-air 
interface and $d=$3.14 mm 
ball are presented in Fig.2. The data taken with increasing and decreasing values of velocities are 
reproducible and show no hysteresis. Analogous data for $d=$2.35 and $d=$1.57 mm balls were also obtained.
Since 
the critical velocity of the transition to the wave generation state is rather low, the critical Reynolds
numbers 
at the transition for different fluids and ball sizes are in the range between 2.2 (for 1.57 mm ball and 
silicone oil 50 cS) and about 700 (for 3.14 mm ball and water). It means that below the transition the drag 
consists of a 
viscous Stokes drag at $Re\leq 1$, which changes linearly with $V$, and the drag, proportional to 
$V^2$, when a boundary layer still remains laminar up to $Re\sim 10^3$\cite{lan}. One fits the data 
below the transition by a second order polynomial and gets a reasonable value of the Stokes drag (see the 
lower inset in Fig.2, where the fit gives $\nu=$8.4 cS instead of 10 cS). Then by subtraction the viscous 
drag one obtains the reduced wave drag force, $(F-F_v)/F_c$, as a 
function of the velocity. This plot clearly exhibits the continuous bifurcation to the wave drag state 
at $V_c=16.5 $ cm/sec, that 
is rather close to the theoretical value of 17.1 cm/sec.
Here $F_v$, $F_c$ are the viscous drag force and the critical value of the full drag force at the onset, 
respectively.
 
 To emphasize a role of a contact line 
relocation due to the ball displacement, two set of data-one with a feedback as in Fig.2, and another
without-are shown in Fig.3 for silicone oil of 50 cS. Since the capillary force exceeds the wave 
drag force near the onset, the relocation of the contact line due to a ball motion and 
 variation of capillary forces in this way diminish partially 
 or completely the result of the wave drag force in the measurements. So, the transition is almost 
 smeared out 
 in the measurements without the feedback. One should mention that in 
our set-up it was impossible to actuate the feedback with a ball less than 1.57 mm dia. Indeed, the viscous 
drag 
reduces with a radius between linear and square dependence but the stabilizing magnetic forces reduces in
a third power. Fig.4 shows the data of the reduced drag force vs the reduced velocity for three
different fluids with 3.14 mm ball: silicone oil(50 cS), glycerol-water(30 cS), and glycerol-water(46 cS). 
It is obvious from the plots that the increase in the wave drag force strongly depends on fluid properties
($\nu$ and $\sigma$). We would like also to point out that the wave drag for water(see the inset in Fig.4)
increases even more dramatically. The inset 
presents the data for water with 3.14 mm ball in the narrower range of the control parameter, 
$V/V_c$, together with a fit by the stationary Ginzburg-Landau equation with a field: 
$\epsilon\xi-a\xi^3+h=o$, where $\xi\equiv(F-F_v)/F_c$ is the order parameter, $\epsilon\equiv V/V_c-1$ is
the control parameter, $a$ is the nonlinear 
coefficient, and $h$ is the field, which plays a role of the smearing factor. As we found from the fits
both $a$ and $h$ depend on $\nu$ and $\sigma$ of fluids, and ball sizes. 
The scaling found experimentally, is $a\sim \eta\sigma/d^2$. Indeed, using this scaling all the 
data for various fluids 
and ball sizes in the vicinity of the transition collapse onto one curve(see Fig.5 for 5 different 
sets of the data). Thus, the transition to the wave drag state is continuous, and the drag force behaves 
above the threshold as $(F-F_v)\sim \sqrt{V-V_c}$.

Fig.6 presents an image of the water surface waves just above the transition that 
 provides information about the opening angle of the 
Cherenkov cone and the wavelength. The inset in Fig.7 shows the dependence of $\lambda$ 
in water on the velocity together with the fit. The main plot presents the data for 
$\cos\theta$ as a function of the velocity. Due to dispersion the phase velocity of the waves depends on the 
wavenumber. We used the fit of the wavenumbers vs velocity  from the inset in Fig.7 to obtain the 
corresponding values of 
$c(k)$ in eq.(3). The result of the calculations of $\cos\theta$ according to eq.(3), is shown by a 
solid line 
in Fig.7. If one takes into account that no fitting parameters were used in this procedure, the agreement 
with the data is rather surprising. Particularly, the critical velocity, $V_c$, obtained from extrapolation
of $\theta\rightarrow 0$, is sufficiently close to the value, found from the force measurements.

Thus, the experiments presented shows unambiguously that the transition to the wave resistance state is
continuous one, contrary to the theoretical prediction\cite{raph}. What can be the reason for this 
discrepancy? The theory is based on the Kelvin model\cite{kelv},
 which is a reasonable approach to long wavelength gravitational waves\cite{our}. Instead of a 
 real ship-like object Kelvin considered a moving  pressure point applied along its course on a water
 surface\cite{kelv}. In a case of an object
 much larger than $\lambda_c$, one can neglect a pressure redistribution due to capillary 
 effects, and the Kelvin model works extremely well for long gravity waves. In the opposite case, 
 a relocation of a contact line 
 causes redistribution of the pressure. That results in a bump behind the moving object observed in  
 experiments. This factor makes the applicability of the Kelvin model questionable. The role of 
 capillarity and wetting  is clearly demonstrated experimentally by switching off the feedback
 control, as we discussed above. Moreover, by using a thin 
 wire as an object one can increase drastically the role of the capillary (wetting) forces in a force 
 balance with the viscous and wave drag forces. Then strong fluctuations of the contact line position 
 due to hysteretic behavior of wetting front, result in strong scatter of the data in the vicinity of 
 the transition which we actually observed in the experiments with wires of 0.3 and 0.7 mm. This effect 
 leads the authors of Ref.\cite{bacri} to an erronous 
 conclusion about the type of the transition to the wave drag state.
 
This experiment was initiated in numerous discussions with a generation of graduate students
in the group. One of us (V.S.) is gratefult to M. Assenheimer, 
H. Davidowitz, A. Groisman, E. Kaplan, and D. Rinberg for valuable and helpful
advices, particularly on initial stages of the research, and to M. Assenheimer and  J. Groshaus for 
participation in early versions of the experiment. This work was supported 
by the Minerva Center for Nonlinear Physics of Complex Systems.

\begin{figure}
\caption {Experimental set-up: FC-channel with a fluid; CYL- cylinder 
to measure a deviation from an initial position by eddy current gauge(EMG); IA- amplifier, and PS-power 
apmlifier; C- coils of electromagnets.}

\label{figa}
\end{figure}
\begin{figure}
\caption {The drag force vs the velocity for silicone oil 
DC200/10 cS and 3.14 mm ball. The upper inset: the same data but in the reduced force.
 The lower inset: viscous drag force 
below the transition vs velocity. Solid line is the second order polynomial fit. }

\label{figb}
\end{figure}
\begin{figure}
\caption {The reduced drag force vs the reduced velocity for a 
silicone oil DC200/50 cS and 3.14 mm ball: full squares-with feedback,
open squares-without feedback.}

\label{figc}
\end{figure}
\begin{figure}
\caption {The reduced drag force vs the reduced velocity for three fluids and 3.14 mm ball: squares-
DC200/50 cS; circles-glycerol-water 30 cS; triangles-glycerol-water 46 cS. The inset: the same for water,
solid curve is the fit by the Ginzburg-Landau equation with a field(see text).}

\label{figd}
\end{figure}
\begin{figure}
\caption {The scaled data of the reduced drag force vs the reduced velocity for 5 different fluids with 
3.14 mm ball except otherwise mentioned: full squares-water; full circles-DC200/50 cS; triangles- 
glycerol-water 46 cS; open squares-glycerol-water 30 cS; open circles-DC200/50 cS, 2.35 mm ball.}

\label{fige}
\end{figure}
\begin{figure}
\caption {Image of the surface waves on water. }

\label{figf}
\end{figure}
\begin{figure}
\caption { $\cos\theta$ of the Cherenkov cone as a function of the velocity for the surface waves on water.
The solid line is calculation based on eq.(1) and (3) with use of the data on $k(V)$, presented in
the inset.}

\label{figg}
\end{figure}

\end{multicols}

\begin{references}
\bibitem{lan} { L. D. Landau and E. M. Lifshitz, {\sl Fluid Mechanics}, 2nd ed
 (Pergamon Press, New York, 1987).}
\bibitem {light}{ J. Lighthill, {\sl Waves in Fluids}, (Cambridge University Press, Cambridge), 1996.}
\bibitem{cher} {  P. A. Cherenkov, C. R. Acad. Sci. USSR {\bf 8}, 451 (1934);
                 I. Frank and I. Tamm, C. R. Acad. Sci. USSR {\bf 14}, 109 (1937).}
\bibitem{lan1} { L. D. Landau and E. M. Lifshitz, {\sl Statistical Physics},
 Part 2, 3rd ed (Pergamon Press, New York, 1980).}
\bibitem{pom} {Y. Pomeau and S. Rica, Phys. Rev. Lett. {\bf 71}, 247 (1993);
 Y. Pomeau, Int. J. of Bifurcation and Chaos in Appl. Sci. and Eng. {\bf 4},
 1165 (1994).}
\bibitem{raph} { E. Rapha$\ddot{e}$l and P.-G. de Gennes, Phys. Rev. {\bf E53}, 3448
 (1996); D. Richard and E. Rapha$\ddot{e}$l, Europhys. Lett. {\bf 48}, 53 (1999).}

\bibitem{kelv} { Lord Kelvin, Proc. Roy. Soc. London {\bf A42} 80 (1887).}


\bibitem {our} { M. I. Shliomis and V. Steinberg, Phys. Rev. Lett. {\bf 79}, 4178 (1997).}
\bibitem{bacri} { J. Browaeys, J.-C. Bacri, C. Flament, S. Neveu, and R. Perzynski,
Eur. Phys. J. B, {\bf 9}, 335 (1999).}

\end{references}
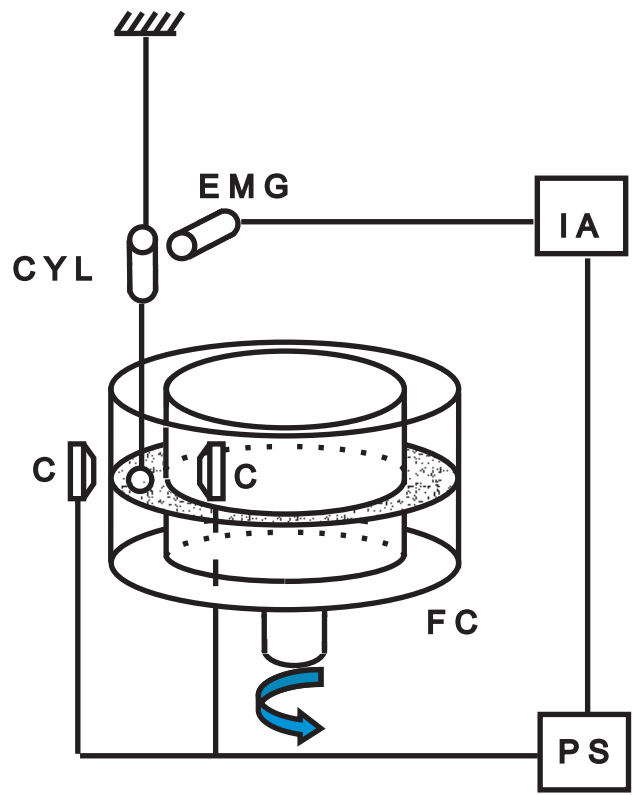
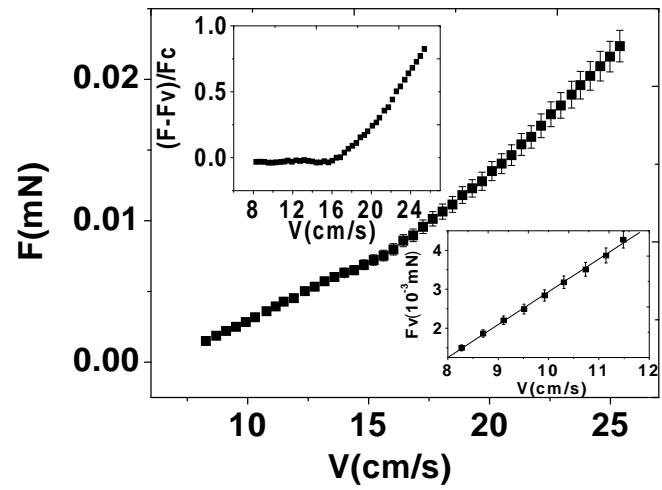
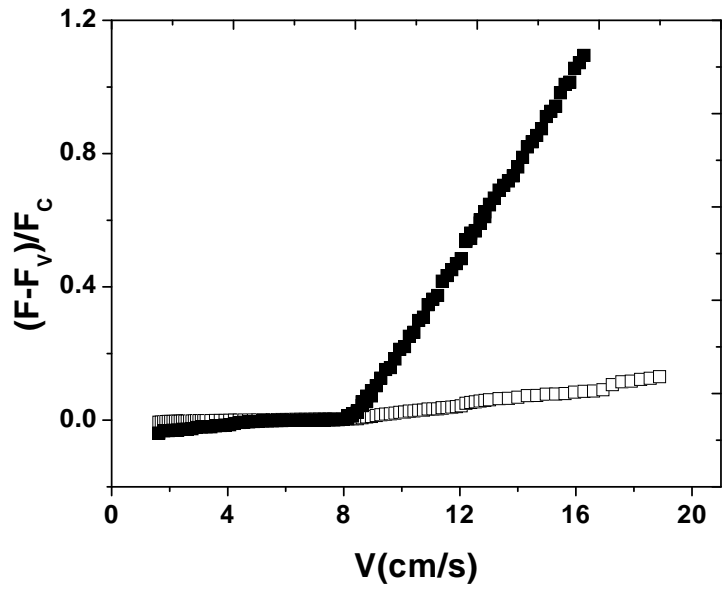
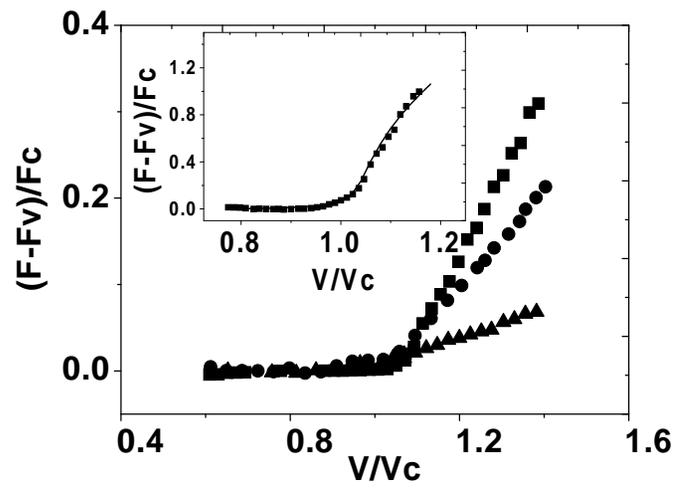
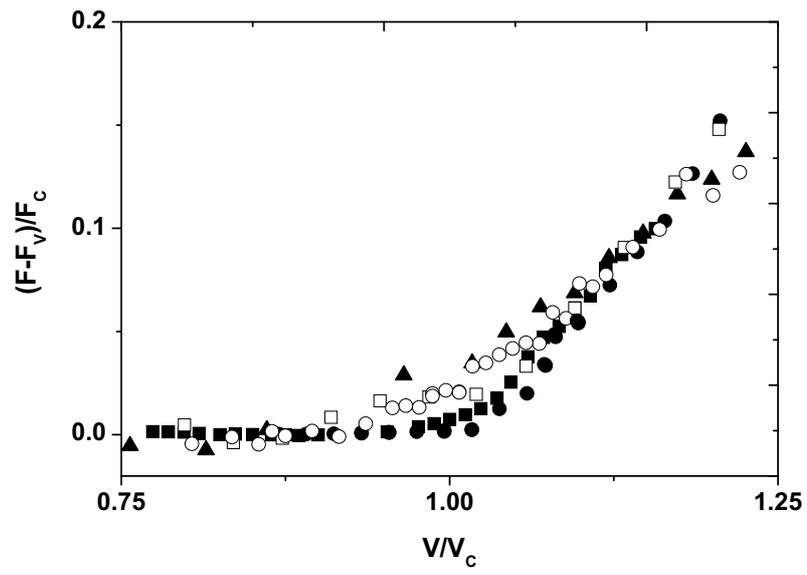
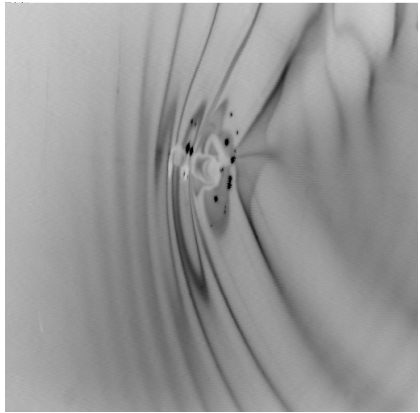
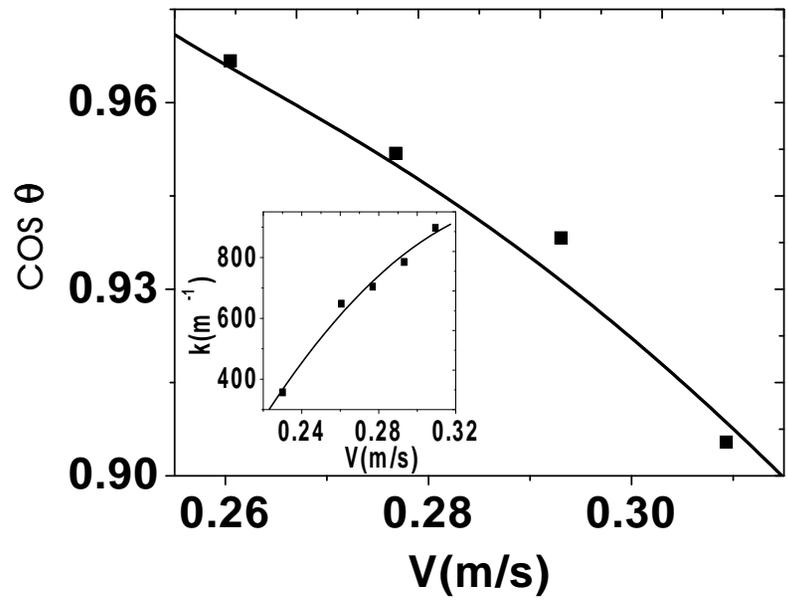
\end{document}